\documentclass[10pt,journal,IEEE]{IEEEtran}
\usepackage{algorithm}
\usepackage{algorithmic}
\usepackage{amsfonts}
\usepackage{amsmath}
\usepackage{bm}
\usepackage[usenames,dvipsnames]{color}
\usepackage[dvips]{graphicx}
\usepackage[update,prepend]{epstopdf}
\usepackage{graphicx}
\usepackage{caption}
\usepackage{subcaption}
\usepackage{makeidx}
\usepackage{subfig}
\usepackage{siunitx}
\usepackage{sidecap}
\usepackage{latexsym}
\usepackage{amssymb}
\usepackage{units}
\usepackage{wasysym}

\ifCLASSOPTIONcompsoc
        \usepackage[nocompress]{cite}

\else

        \usepackage{cite}

\fi

\newcommand{\vect}[1]{\mathbf{#1}}

\newcommand{\kHz}{\text{kHz}}

\title{Frequency Selective Compressed Sensing }

\author{Jacek Pierzchlewski, Thomas Arildsen\\
Aalborg University\\
Faculty of Engineering and Science \\
Department of Electronic Systems \\
Fredrik Bajers Vej 7B, 9220 Aalborg \O{}, Denmark\\
jap@es.aau.dk, tha@es.aau.dk\\
}

\begin{document}

\onecolumn
\Huge
This work has been submitted to the IEEE for possible publication.
Copyright may be transferred without notice, after which this version may no longer be accessible

\normalsize
\twocolumn
\maketitle
\thispagestyle{empty}

\begin{abstract}
In this paper the authors describe the problem
of acquisition of interfered signals and formulate
a filtering problem.
A frequency-selective compressed sensing technique
is proposed as a solution to this problem.
Signal acquisition is critical in facilitating frequency-selective compressed sensing.
The authors propose a filtering compressed sensing parameter, which
allows to assess if a given acquisition process makes
frequency-selective compressed sensing possible for a given filtering problem.
A numerical experiment which shows how the described method works
in practice is conducted.
\end{abstract}

\begin{IEEEkeywords}
        Analog-digital conversion, Compressed sensing, Signal sampling,
        Interference reduction,
\end{IEEEkeywords}

\section{Introduction}

The
Shannon-Nyquist sampling theorem states that perfect signal reconstruction
of any signal requires a sampling frequency higher than twice the maximum
frequency component in the signal \cite{pat_Nyq01}.
In practical situations we thus need analog anti-aliasing filters prior
to the analog-to-digital conversion (ADC) to facilitate the above
\cite{pat_Lyons10, pat_Mit95, pat_An01, pat_Le05}
and to reduce the risk of saturating the ADC due to limited dynamic range,
causing nonlinear distortion
\cite{pat_An01, pat_Le05, pat_Bes03}.
However, applying such analog filtering is design and implementation
challenging \cite{pat_Bah12, pat_Raz98, pat_Pat12, pat_Yeu12},
particularly in the radio frequency range.
Therefore a digital solution would be preferred if possible.

In this paper the authors propose a frequency-selective sampling
method based on the compressed sensing technique.
In recent years a new idea of signal sensing,
known as compressed sensing (CS) has emerged
\cite{pat_Cand01, pat_Ela01, pat_Bar01, pat_Cand02, pat_Don01}.
This technique can be used to successfully
reconstruct signals that are sampled at a sub-Nyquist rate,
provided the signal is sparse in some domain.
This technique is well elaborated, however,
to the authors' best knowledge there are no publications
in which compressed sensing reconstruction selectively favors certain signal
spectrum frequencies over others.

Frequency-selective compressed sensing proposed in this paper allows for
wider frequency spectrum of the sampled signal then just the wanted signal.
Hence, relaxed input signal filtering is possible
without increasing the sampling frequency to the Nyquist
frequency of the input signal polluted by unwanted high-frequency signals.
The main idea is to divide the compressed sensing reconstruction procedure into
two phases: an optimization phase and a final signal reconstruction phase,
in the latter phase a limited signal dictionary is used.
The authors propose a filtering compressed sensing (fCS) parameter,
which assesses
if a used acquisition process makes
frequency-selective compressed sensing possible.

The paper is organized as follows.
The problem considered in this paper is discussed in Section \ref{sec:problem}.
Frequency-selective compressed sensing
is described in Section \ref{sec:fCS}.
Filtering compressed sensing (fCS) parameter
is proposed in the Section \ref{sec:safp}.
The performance of the proposed solution is assessed by numerical simulation
in Section \ref{sec:exper}.
The paper is concluded in Section \ref{sec:conc}.
The paper follows the reproducible research paradigm \cite{pat_Vand01},
all the code and figures associated with the experiment are available online \footnote{Aalborg University (2014). ``IRfDUCS project'', [Online] Available: http://www.irfducs.org/fRIP}.

\section{Problem formulation}
\label{sec:problem}

        \begin{figure}[b]
                        \centering
                        \includegraphics{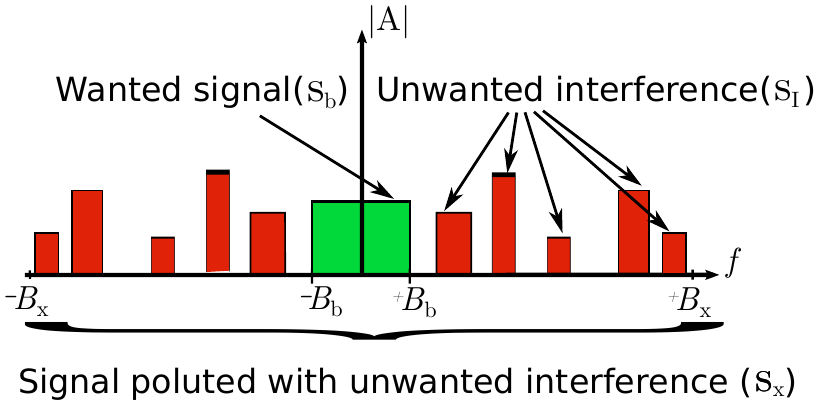}
                        \caption{Frequency spectrum of the received signal $\vect{s}_{\text{x}}$. The signal consists of
                                the wanted signal $\vect{s}_{\text{b}}$ (green) and the unwanted interference $\vect{s}_{\text{I}}$ (red).
                        The signal's frequency range is ($-B_{\text{x}} < f \leq B_{\text{x}}$).}
                        \label{fig:spectrum_original}
        \end{figure}

        \begin{figure}[t]
                \centering
                        \includegraphics{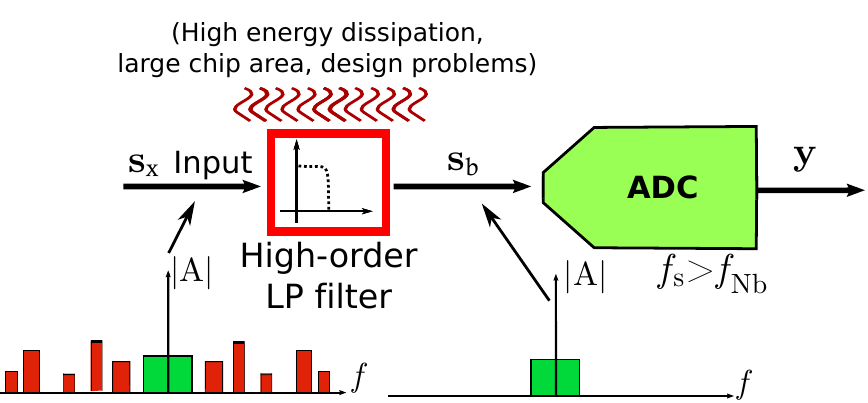}
                        \caption{Solution 1: High-order filtering enables lowered sampling frequency.
                        Large filters utilize considerable chip area and cause high energy dissipation and integrated circuit design problems.}
                        \label{fig:case1_high_filtering_low_samp_freq}
        \end{figure}

        Let us define a bandlimited ($-B_{\text{b}} \leq f \leq B_{\text{b}}$) baseband signal
        $\vect{s}_{\text{b}} \in \mathbb{R}^{M \times 1}$.
        %The signal $\vect{s}_{\text{b}}$ contains information and is received and processed by a given digital processing system.
        The Nyquist rate of the signal $\vect{s}_{\text{b}}$ is $f_{\text{Nb}} = 2B_{\text{b}}$.
        The signal is polluted by an interference passband $( B_{\text{b}} < f \leq B_{\text{x}}, \; -B_{\text{x}} \leq f < -B_{\text{b}} )$
        signal $\vect{s}_{\text{I}} \in \mathbb{R}^{M \times 1}$.
        The received signal $\vect{s}_{\text{x}} \in \mathbb{R}^{M \times 1}$ is a sum of
        the wanted signal $\vect{s}_{\text{b}}$ and the interference signal $\vect{s}_{\text{I}}$:
        $\vect{s}_{\text{x}} = \vect{s}_{\text{b}} + \vect{s}_{\text{I}}$ (Fig. \ref{fig:spectrum_original}).
        The signal $\vect{s}_{\text{x}}$ is bandlimited ($-B_{\text{x}} \leq f \leq B_{\text{x}}$),
        its Nyquist rate is $f_{\text{Nx}} = 2B_{\text{x}}$.

        \begin{figure}[h!]
                \centering
                        \includegraphics{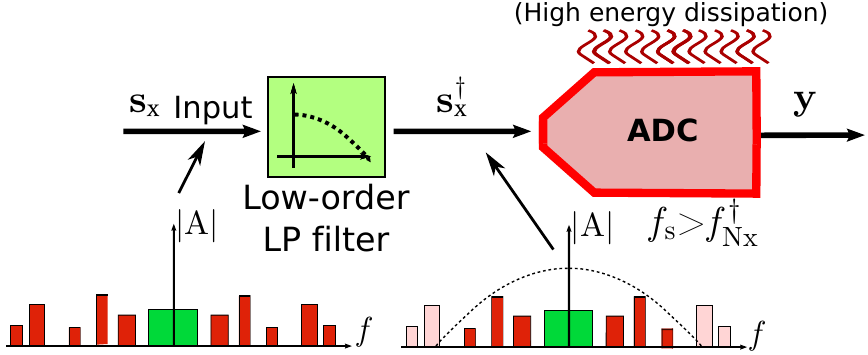}
                        \caption{Solution 2: Low-order filtering requires high sampling frequency.
                                 High sampling frequency causes high energy dissipation and is infeasible to
                                be implemented in certain applications.}
                        \label{fig:case2_low_filtering_high_samp_freq}
        \end{figure}

        \begin{figure}[h!]
                        \centering
                        \includegraphics{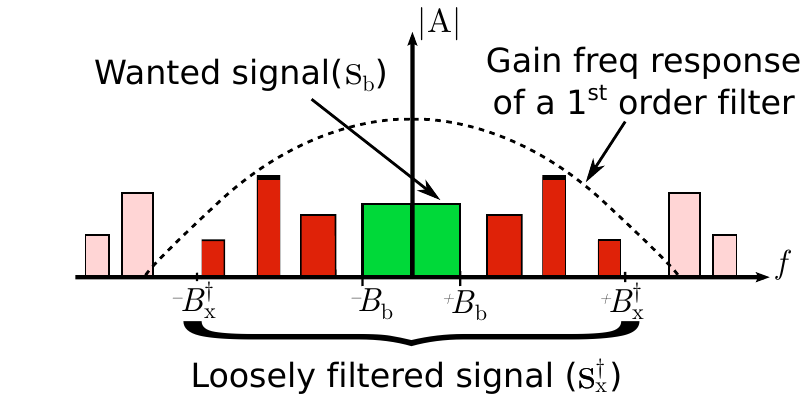}
                        \caption{Frequency spectrum of the loosely filtered signal $\vect{s}^{\dag}_{\text{x}}$ which
                                consists of the wanted signal $\vect{s}_{\text{b}}$ (green) and the loosely filtered interference signal (red)
                        which is still partly present in the spectrum. A part of the unwanted frequencies was removed (pale red).}
                        \label{fig:spectrum_loosely_filtered}
        \end{figure}

        Due to the interference signal $\vect{s}_{\text{I}}$,
        the Nyquist rate $f_{\text Nx}$ of the received signal $\vect{s}_{\text x}$
        is in many applications significantly higher than the Nyquist rate $f_{\text{Nb}}$
        of the wanted baseband signal $\vect{s}_{\text{b}}$.
        To enable sampling with low frequency
        $f_{\text{s}}$
        ($f_{\text{Nb}} < f_{\text s} \ll f_{\text{Nx}}$) the signal
        $\vect{s}_{\text{x}}$ must be filtered with a high-order
        low-pass filter which removes the unwanted interference
        (Fig. \ref{fig:case1_high_filtering_low_samp_freq}).
        Unfortunately, high-order filters cause design and
        integrated circuit implementation problems due to
        high energy dissipation and chip area required
        to implement these filters \cite{pat_Raz98,pat_Pat12,pat_Yeu12}.

        Another possibility is to "loosely" filter a signal with a low-order
        filter (Fig. \ref{fig:case2_low_filtering_high_samp_freq}).
        Let us consider a bandlimited ($-B^{\dag}_{\text{x}} \leq f \leq B^{\dag}_{\text{x}}$)
        signal $\vect{s}^{\dag}_{\text{x}}$, which is created by applying a
        1$^{\text{st}}$-order filter on the
        received signal $\vect{s}_{\text{x}}$.
        This partly removes high-frequency unwanted signals,
        however there is still considerable interference content present
        in the filtered signal $\vect{s}^{\dag}_{\text{x}}$ (Fig. \ref{fig:spectrum_loosely_filtered}).
        The Nyquist frequency of the signal $\vect{s}^{\dag}_{\text{x}}$
        is $f^{\dag}_{\text{Nx}} = 2B^{\dag}_{\text{x}}$.
        The baseband $B^{\dag}_{\text{x}}$ of the filtered signal depends on the
        filter's cut-off frequency $f_{\text{c}}$.
        The Nyquist rate $f^{\dag}_{\text{Nx}}$ of the filtered signal $s^{\dag}_{\text{x}}$
        is lower than the Nyquist rate $f_{\text{Nx}}$ of the unfiltered signal $s_{\text{x}}$,
        but in many applications
        it is still significantly higher than the Nyquist rate $f_{\text{Nb}}$
        of just the wanted signal $s_{\text{b}}$:
        \begin{equation}
                f_{\text{Nb}} \ll f^{\dag}_{\text{Nx}} \ll f_{\text{Nx}}
        \end{equation}
        Therefore, if a low-order filter is used, a high sampling frequency
        must be applied to the signal (Fig. \ref{fig:case2_low_filtering_high_samp_freq}).
        The high sampling frequency causes high energy dissipation
        \cite{pat_Le05,pat_Bes03,pat_Bah12}
        and may be infeasible to implement in certain applications.

%%% Local Variables:
%%% mode: latex
%%% TeX-master: "paperfCS"
%%% End:

\section{Frequency-Selective Compressed Sensing}
\label{sec:fCS}
Compressed sensing is a technique which allows for
signal sampling with frequency lower then the signal's Nyquist rate.
Compressed sensing is possible if the sampled signal $\vect{x} \in \mathbb{R}^{M \times 1}$ can be
represented as:
$\vect{x} = \vect{\Psi} \vect{v}$ where
$\vect{\Psi} \in \mathbb{R}^{M \times 2K}$ is a signal's dictionary, $\vect{v} \in \mathbb{R}^{2K \times 1}$
is a sparse vector -- a vector with only few ($\text{S}$) non-zero elements.
The number of non-zero elements ($\text{S}$) is often called the signal's `sparsity'.
A relation between signal sparsity and compressed sensing is well
developed in \cite{pat_Cand02}.

Compressed sensing can be divided into two parts:
signal acquisition and signal reconstruction.
An observed signal $\vect{y} \in \mathbb{R}^{N \times 1}$ is an outcome of the acquisition process: $\vect{y} = \vect{\Phi} \vect{x} $,
where the sensing matrix $\vect{\Phi} \in \mathbb{R}^{N \times M}$ represents the (linear) acquisition process.
The reconstructed signal $\hat{\vect{x}}$ is computed as: $ \hat{\vect{x}} = \vect{\Psi} \hat{\vect{v}} $,
where $\hat{\vect{v}}$ is the reconstructed sparse vector.
There are several methods for reconstructing the sparse vector $\hat{\vect{v}}$.
One of the most classic is basis pursuit denoising or LASSO \cite{LASSO96,BPDN98},
which is an $\ell_1$ optimization process.
This convex optimization problem can be posed as:
\begin{equation}
\label{eq:closed}
\hat{\vect{v}} = \underset{\vect{v}}{\operatorname{argmin}} \|\vect{v}\|_1 + k \| \vect{y} - \vect{\Theta}\vect{v}\|_2^2
\end{equation}

        The authors' aim is to decrease the necessary sampling frequency in case
        the received signal is $\vect{s}_{\text{x}}^{\dag}$ with
        bandwidth $B^{\dag}_{\text{x}}$,
        while only a lower-frequency
        part of the signal
        needs
        to be correctly reconstructed (Fig. \ref{fig:spectrum_loosely_filtered}).
        The filtering problem $\mathcal{P}(B_{\text x}^{\dag},B_{\text b}, \vect{\Psi}, \text{S})$
        is constituted by four parameters:
        baseband of the interfered signal $B_{\text x}^{\dag}$,
        wanted signal baseband $B_{\text b}$,
        dictionary $\vect{\Psi}$, and signal sparsity $\text{S}$.

The dictionary matrix $\vect{\Psi}$ used in this paper is a Digital Hartley Transform (DHT) matrix \cite{Sorensen01}.
The matrix consists of $2K$ columns indexed $k \in \{1, ..., 2K\}$.
Frequencies reflected by the columns of the dictionary matrix are,
for the first $K$ columns:
$f^{k}_{\text{low}} = (-K + k -1)\delta_{f}, k \in \{1,...,K\}$,
for the last $K$ columns:
$f^{k}_{\text{high}} = (-K + k)\delta_{f}, k \in \{K+1,...,2K\}$,
where $\delta_{f}$ is the frequency separation between the dictionary columns.
The dictionary matrix used must span the full spectrum
of the loosely sampled signal $\vect{s}_{\text{x}}^{\dag}$.
In a typical compressed sensing problem there is a need to reconstruct
all the coefficients in the vector $\hat{\vect{v}}$ correctly,
but, as described in Section \ref{sec:problem}, in the current problem
it is only necessary to reconstruct the frequencies corresponding
to the wanted signal $\vect{s}_{\text{b}}$.
Indices $k'_{\text{b}} \in \vect{K'_{\vect{b}}}$ of the columns of the
dictionary $\vect{\Psi}$
which correspond to the signal $\vect{s}_{\text{b}}$ are
within the interval:
\begin{equation}
\label{eq:interval}
\vect{K'_{\vect{b}}} = \{ K - \alpha + 1,..., K + \alpha \} \; \; \;\; \alpha = \lceil {B_{\text{b}}} / \delta_{f} \rceil
\end{equation}
The wanted reconstructed signal $\hat{\vect{s}}_{\text{b}}$
can be computed as:
\begin{equation}
\label{eq:vectorv}
\hat{\vect{s}}_{\text{b}} = \vect{\Psi}_{\text{b}} \hat{\vect{v}}_{\text{b}}
\end{equation}
\begin{figure}[t]
		\centering
		\includegraphics{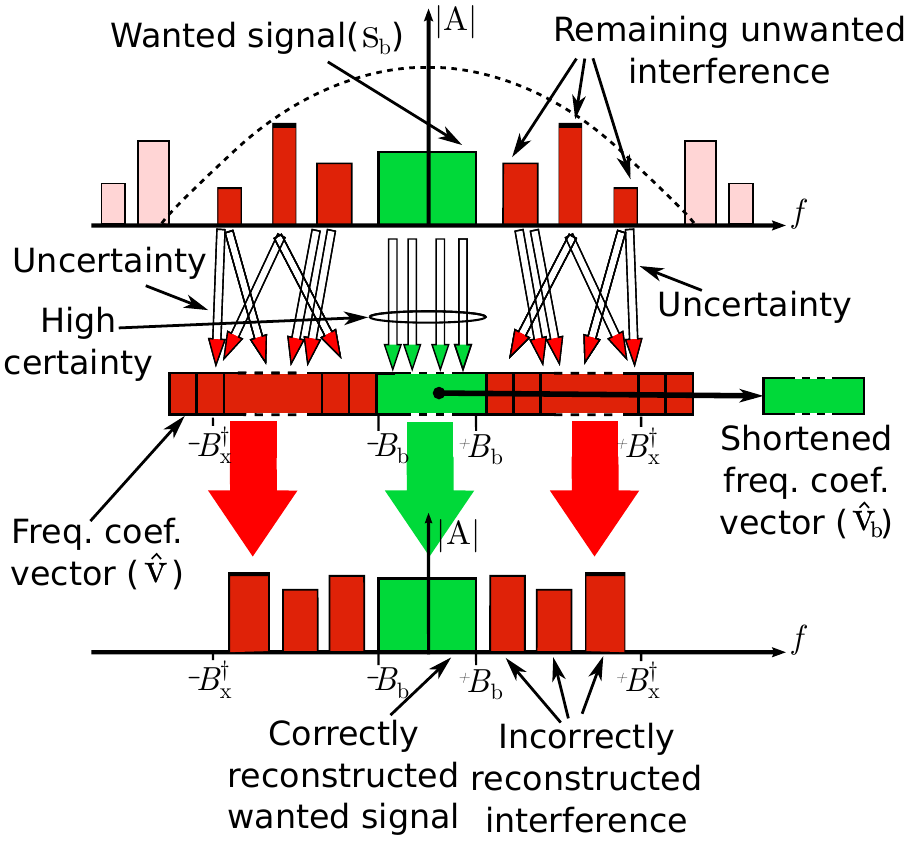}
		\caption{The idea of frequency-selective compressed sensing}
		\label{fig:cert}
\end{figure}
where $\vect{\Psi}_{\text{b}}$ is a dictionary
composed of columns
$k'_{\text{b}} \in \vect{K'_{\vect{b}}}$ within the inverval as in (\ref{eq:interval}).
The vector $\hat{\vect{v}}_{\text{b}} = \hat{\vect{v}}[k'_{\text{b}}]$
is the central part of the reconstructed vector $\hat{\vect{v}}$
composed by the elements of $\hat{\vect{v}}$
corresponding to the frequency spectrum of the signal
$\vect{s}_{\text{b}}$.
It is clear that the reconstruction process (\ref{eq:closed})
only needs to reconstruct the
$\hat{\vect{v}}_{\text{b}}$ part of the vector $\hat{\vect{v}}$ correctly.
Therefore, the acquisition process must be tailored so that it
brings certainty into the reconstruction of the $\hat{\vect{v}}_{\text{b}}$ part,
while uncertainty in the reconstruction of the rest of the vector $\hat{\vect{v}}$ is
allowed (Fig. \ref{fig:cert}).
Signal which is not covered by the dictionary is seen by a compressed sensing reconstrucion process as noise.
Therefore, a dictionary $\vect{\Psi}$ which spans all the spectrum of the received signal $\vect{s}^{\dag}_{\text{x}}$
must be used in the reconstruction process.
Otherwise, the interfering signal would be treated by the reconstruction algorithm
as noise in the sampled signal, which would dramatically compromise the quality of the reconstruction process.

%%% Local Variables:
%%% mode: latex
%%% TeX-master: "paperfCS"
%%% End:

\section{Signal Acquisition for Filtering Problems}
\label{sec:safp}
       \begin{figure}[b!]
                        \centering
                        \includegraphics{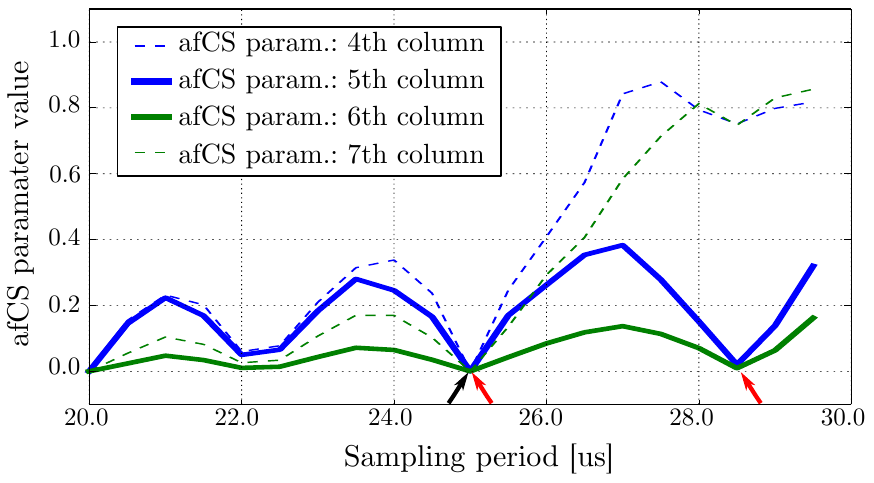}
                        \caption{Atomic filtering compressed sensing (afCS) parameters for columns 4 - 7 of the dictionary matrix. The parameters were computed for different sampling periods.}
                        \label{fig:afCS}
        \end{figure}

Designing the signal acquistion part of a compressed sensing process
is critical in facilitating frequency-selective compressed sensing.
In this section a filtering CS parameter is introduced.
It can be used to evaluate if a given acquisition process,
represented by a sensing matrix $\vect{\Phi}$,
is suitable for a filtering problem $\mathcal{P}$.
\subsection{Evaluation of Signal Acquisition for Frequency-Selective Compressed Sensing}
Let us define a matrix $\vect{\Theta} \in \mathbb{R}^{\text{N} \times 2\text{K}}$
which consists of normalized columns of the matrix
$\vect{\Theta}' \in \mathbb{R}^{\text{N} \times 2\text{K}}$:
\begin{equation}
\vect{\Theta}' = \vect{\Phi} \vect{\Psi}
\end{equation}
where
$\vect{\Psi} \in \mathbb{R}^{\text{M} \times 2\text{K} }$ is a dictionary matrix,
and $\vect{\Phi} \in \mathbb{R}^{\text{N} \times \text{M}}$ is a sensing matrix
which represents the compressed sensing acquisition process.
Hence, it can be stated that $\vect{\Theta} = f_{\text{n}}(\vect{\Phi},\vect{\Psi})$, where
$f_{\text{n}}$ is a column-wise normalization function.
Let us define an atomic filtering compressed sensing (afCS) parameter:
\begin{equation}
\label{eq:atomic}
\zeta_{k}^{\vect{\Theta}} = f_{\text{f}}(\vect{\Theta},k),
\end{equation}
the function $f_{\text{f}}$ is defined in Section~\ref{sub:atomic}.
The parameter $\zeta_{k}^{\vect{\Theta}}$ signifies how well the $k$\textsuperscript{th} entry of the sparse vector $\hat{\vect{v}}$ (\ref{eq:vectorv})
will be reconstructed by a reconstruction algorithm for a given matrix $\vect{\Theta} = f_{\text{n}}(\vect{\Phi},\vect{\Psi})$ --
the lower the parameter $\zeta_{k}^{\vect{\Theta}}$, the better
the reconstruction of the $k$\textsuperscript{th} entry.
Now let us define a filtering compressed sensing parameter (fCS) $\xi^{\vect{\Theta}}_{\vect{B}_{\text{b}}}$:
\begin{equation}
\label{eq:filtCSpar}
\xi^{\vect{\Theta}}_{\vect{B}_{\text{b}}} = \smash{\displaystyle\max_{k \in \vect{K'_{\vect{b}}}}} (\zeta_{k}^{\vect{\Theta}})
\end{equation}
where $\vect{K'_{\vect{b}}}$ defined as in (\ref{eq:interval}),
is the set of indices of columns of the $\vect{\Theta}$ matrix
corresponding to the wanted signal $\vect{s}_{\text{b}}$.
The filtering compressed sensing parameter $\xi^{\vect{\Theta}}_{\vect{B}_{\text{b}}}$ is calculated for a set of columns of $\vect{\Theta}$,
while the atomic filtering compressed sensing parameter $\zeta_{k}^{\vect{\Theta}}$ is calculated for a single column of $\vect{\Theta}$.
To facilitate frequency-selective compressed sensing for a given filtering problem $\mathcal{P}(B_{\text x}^{\dag},B_{\text b}, \vect{\Psi}, \text{S})$,
one must find a sensing matrix $\vect{\Phi}$ for which the filtering compressed sensing (fCS) parameter
$\xi^{\vect{\Theta} = f_{\text{n}}(\vect{\Phi},\vect{\Psi}) }_{\vect{B}_{\text{b}}}$ is close to zero.

\subsection{Computation of The Atomic Filtering Compressed Sensing Parameter}
\label{sub:atomic}
Here the authors show how to realize the function $f_{\text{f}}$ from (\ref{eq:atomic}) which
computes an atomic filtering compressed sensing parameter $\zeta_{k}^{\vect{\Theta}}$
for the $k$\textsuperscript{th} column of the matrix $\vect{\Theta}$.

For $k$\textsuperscript{th} column $\theta_{k}$ of the matrix $\vect{\Theta} $ let us create a projection matrix $\vect{P}_{k}$:
\begin{equation}
\vect{P}_{k} = \vect{\theta}_{k} (\vect{\theta}_{k}^{\text{T}} \vect{\theta}_{k} )^{-1} \vect{\theta}^{\text{T}}_{k}
\end{equation}
Let us generate a matrix $\vect{\Omega}$, which contains $\text{W}$
testing vectors as columns.
The matrix $\vect{\Omega} \in \mathbb{R}^{\text{2K} \times \text{W}} $
is composed of normalized columns from the
matrix $\vect{\Omega}' \in \mathbb{R}^{\text{2K} \times \text{W}}$
the elements $\omega'_{k, w}$ of which are random Gaussian values:
\begin{equation}
\omega'_{k,w} = \vect{\Omega}'(k, w), \; \omega'_{k,w} \sim \mathcal{N}(0, 1)
\end{equation}
Let us define the matrix $\vect{A} \in \mathbb{R}^{\text{N} \times \text{W}}$
which is the product of the matrix $\vect{\Theta}$ by the matrix of testing vectors $\vect{\Omega}$:
\begin{equation}
\vect{A} = \vect{\Theta} \vect{\Omega}
\end{equation}
Now it is possible to compute a matrix $\vect{\Gamma}_{k} \in \mathbb{R}^{\text{N} \times \text{W}}$
which contains vectors from the matrix $\vect{A}$
projected onto the $k$\textsuperscript{th} column of the matrix $\vect{\Theta}$:
\begin{equation}
\vect{\Gamma}_{k} = \vect{P}_{k} \vect{A}
\end{equation}
The atomic filtering compressed sensing (afCS) parameter $\zeta_{k, w}^{\vect{\Theta}}$ of the $k$\textsuperscript{th} column,
computed for the $w$\textsuperscript{th} testing vector is:
\begin{equation}
\zeta^{\vect{\Theta}}_{k, w} = | \| \gamma^{w}_{k} \|_{2} - \vect{\Omega}(k, w )   |
\end{equation}
where $\gamma^{w}_{k}$ is the $w$\textsuperscript{th} column of the matrix $\vect{\Gamma}_{k}$.
An estimated atomic filtering compressed sensing parameter
of the $k$\textsuperscript{th} column of the matrix $\vect{\Theta}$ is:
\begin{equation}
\zeta^{\vect{\Theta}}_{k} = \underset{w \in \text{W}}{\max}(\zeta^{\vect{\Theta}}_{k, w})
\end{equation}
It requires an infinite number ($\text{W}\to\infty$) of testing wectors to determine the correct value of
$\zeta^{\vect{\Theta}}_{k}$ numerically.
In practice, the number of testing vectors
needed to determine
$\zeta^{\vect{\Theta}}_{k}$ with sufficient accuracy
should be found experimentally.

%%% Local Variables:
%%% mode: latex
%%% TeX-master: "paperfCS"
%%% End:

\section{Numerical Experiment}
\label{sec:exper}
A numerical experiment was conducted to verify the idea practically.
Loosely filtered signal $\vect{s}_{\text{x}}^{\dag}$ (Fig. 2) consists of maximum 5 tones
separated by 5 \kHz, its Nyquist frequency is 50 \kHz.
The total number of tones currently present in the signal is not known to the recontruction algorithm.
The signal dictionary $\vect{\Psi}$ used in the experiment is a discrete
Hartley transform dictionary
with 10 columns which reflect 5 frequencies: $\{5 \; \kHz, 10 \; \kHz, ..., 25 \; \kHz \}$.
Let us define two filtering problems $\mathcal{P}_{1}(25 \kHz, 5 \kHz, \vect{\Psi}, N_{\text{b}}+N_{\text{I}})$
and $\mathcal{P}_{2}(25 \kHz, 10 \kHz, \vect{\Psi}, N_{\text{b}}+N_{\text{I}})$, where
$N_{\text{b}}$ is the number of wanted tones,
$N_{\text{I}}$ is the number of interfering tones.
In the first problem the lowest frequency tone (5 \kHz) must be correctly reconstructed ($N_{\text{b}} = 1$).
In the second problem two frequency tones (5 \kHz \, and 10 \kHz) must be correctly reconstructed ($N_{\text{b}} = 2$).
The frequency location of the $N_{\text{b}}$ wanted tones is known,
while the frequency location and the number of $N_{\text{I}}$ interfering tones is not known by the reconstruction algorithm.

The signal is uniformly sampled, to check which sampling period is the best
for the filtering problems $\mathcal{P}_{1}$ and $\mathcal{P}_{2}$. Atomic filtering compressed sensing (afCS) parameters
(\ref{eq:atomic}) for columns
$4 - 7$ of the dictionary was measured for different sampling periods.
The sampling period was swept from $20$ $\mu$s to $30$ $\mu$s with $0.5$ $\mu$s step.
The Nyquist frequency of the signal (50 \kHz) corresponds to $20$ $\mu$s sampling period,
so all of the tested uniform sampling frequencies were not higher than
the Nyquist frequency of the loosely filtered signal $\vect{s}_{\text{x}}^{\dag}$.
Measured parameters are plotted in Fig.~\ref{fig:afCS}.
Columns 4 and 7 correspond to a 10 \kHz \, frequency tone,
columns 5 and 6 correspond to a 5 \kHz \, frequency tone.
Filtering compressed sensing parameter $\xi^{\vect{\Theta}}_{\vect{B}_{\text{b}}}$ (\ref{eq:filtCSpar}) for the first filtering problem $\mathcal{P}_{1}$
is computed using atomic filtering compressed sensing parameters for columns 5 and 6:
\begin{equation}
\xi^{\vect{\Theta}}_{5 \kHz} = \max (\zeta_{5}^{\vect{\Theta}} , \zeta_{6}^{\vect{\Theta}})
\end{equation}
while the the parameter computed for the second filtering problem $\mathcal{P}_{2}$
is computed using atomic filtering compressed sensing parameters for columns $\{4,..,7\}$:
\begin{equation}
\xi^{\vect{\Theta}}_{10 \kHz} = \max (\zeta_{4}^{\vect{\Theta}} , \zeta_{5}^{\vect{\Theta}} , \zeta_{6}^{\vect{\Theta}} ,  \zeta_{7}^{\vect{\Theta}})
\end{equation}
The parameter $\xi^{\vect{\Theta}}_{5 \kHz}$ is close to 0 for sampling
periods 25.0 $\mu$s and 28.5 $\mu$s (marked with red arrows in Fig.~\ref{fig:afCS}).
The parameter $\xi^{\vect{\Theta}}_{10 \kHz}$ is close to 0 for the
sampling period of 25 $\mu$s (marked with a black arrow in Fig.~\ref{fig:afCS}).

\begin{figure}[h!]
                        \centering
                        \includegraphics{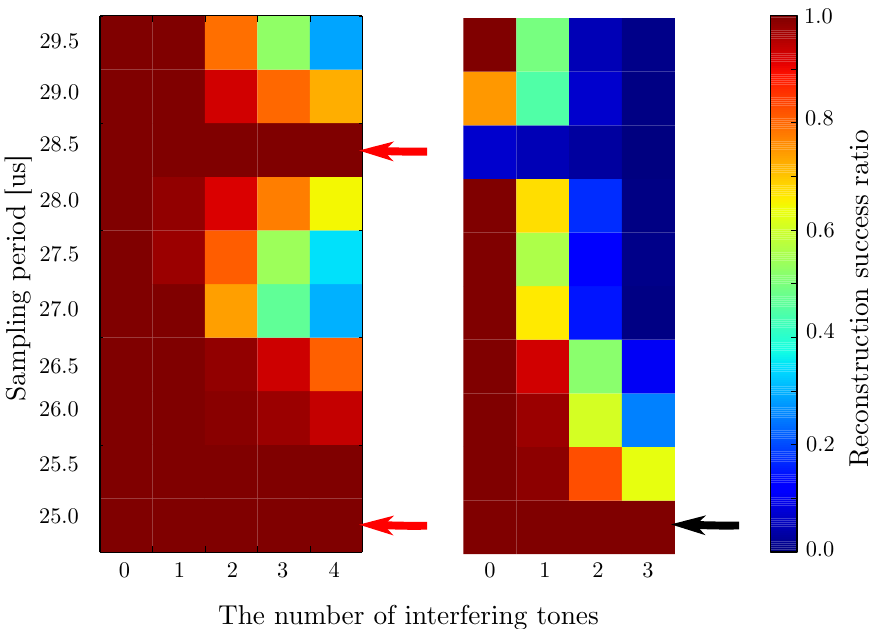}
                        \caption{Reconstruction success ratio of
                        filtering problem $\mathcal{P}_{1}$ (left)
                        and $\mathcal{P}_{2}$ (right)}
                        \label{fig:map1}
        \end{figure}

Average reconstruction success ratio was measured for filtering problems
$\mathcal{P}_{1}$ and $\mathcal{P}_{2}$. The ratio was measured over $10^{4}$ random cases.
The sampling period is swept from $25$ $\mu$s to $30$ $\mu$s with $0.5$ $\mu$s step.
The number $N_{\text{I}}$ of interfering tones is swept over $N_{\text{I}} \in \{0, 1, ..., 4\}$ for
the filtering problem $\mathcal{P}_{1}$ and $N_{\text{I}} \in \{0, 1, ..., 3\}$
for the problem $\mathcal{P}_{2}$.
The reconstruction is treated as successful
if the signal-to-noise ratio of the reconstructed wanted signal
$\vect{s}_{\text{b}}$ is equal to or higher than 25 dB.
The results are plotted in Fig. \ref{fig:map1}.

As expected,
the reconstruction of the wanted signal from
the filtering problem $\mathcal{P}_{1}$
for the sampling periods 25.0 $\mu$s and 28.5 $\mu$s (red arrows) is ideal.
Reconstruction of the wanted signal from
the filtering problem $\mathcal{P}_{2}$
is ideal for the sampling period 25.0 $\mu$s (black arrow).
Surprisingly, no sparsity of the received signal is needed,
since the reconstruction works well when the signal and interference
occupy the whole spectrum.
Instead of signal sparsity, the proposed method exploits
that only a part of the spectrum is required to be reconstructed correctly.

%%% Local Variables:
%%% mode: latex
%%% TeX-master: "paperfCS"
%%% End:

\section{Conclusions}
\label{sec:conc}
In this paper the authors have described the problem
of acquistion of interfered signals and formulated
a filtering problem $\mathcal{P}$.
A frequency-selective compressed sensing technique
was proposed as a solution.
A filtering compressed sensing parameter was proposed for assessing if a given
signal acquisition process makes
frequency-selective compressed sensing possible for a given filtering problem.
A numerical experiment which shows how the method works
in practice was conducted.
%Furthermore, this method does not require complicated acquistion
%methods, and works well with converters dedicated to
%standard signal sensing techniques.

%%% Local Variables: 
%%% mode: latex
%%% TeX-master: "paperfCS"
%%% End: 

\end{document}